\documentstyle[12pt]{article}

\newcommand{\be}{\begin{equation}}
\newcommand{\ee}{\end{equation}}

\newcommand{\bea}{\begin{eqnarray}}
\newcommand{\eea}{\end{eqnarray}}

\newcommand{\bean}{\begin{eqnarray*}}
\newcommand{\eean}{\end{eqnarray*}}

\newcommand{\ba}{\begin{array}}
\newcommand{\ea}{\end{array}}

\newcommand{\g}{\gamma}

\newcommand{\ssu}{$SU(2)_L\times SU(2)_R\times U(1)_{B-L}\,$}
\newcommand{\vev}[1]{\langle{#1}\rangle}
\newcommand{\lsim}{{\;\raise0.3ex\hbox{$<$\kern-0.75em\raise-1.1ex\hbox{$\sim$}}
\;}}
\newcommand{\gsim}{{\;\raise0.3ex\hbox{$>$\kern-0.75em\raise-1.1ex\hbox{$\sim$}}
\;}}

\begin{document}

\begin{titlepage}

\mbox{}\vspace*{-1cm}\hspace*{8cm}\makebox[7cm][r]{\large
HU-SEFT-1995-14}
\vfill

\LARGE

\begin{center} {The $W_R$ production in $e\gamma$ collisions}

\bigskip
\large {K. Huitu \\Research Institute for High Energy Physics,
\\University of Helsinki
\\[15pt] J. Maalampi \\Department of Theoretical Physics,University of
Helsinki
\\[15pt] M. Raidal\footnote{On leave of absence from KBFI, Academy of
Estonia; Address after October 1, 1995: Department of Theoretical
Physics, University of Valencia, Spain}
\\Department of Theoretical Physics,University of Helsinki }
\date{August 1995}

\bigskip

\today

\vfill

\normalsize

{\bf\normalsize \bf Abstract} \end{center}

\normalsize

We consider the production of a single right-handed
gauge boson $W_R$ in the high-energy $e^-\gamma$ collisions with
polarized beams. If the associated neutrino is light, the reaction
will give the best discovery reach for $W_R$ in the Next Linear
Collider.
\bigskip
\end{titlepage}

\newpage

The left-right symmetric model (LRM) \cite{lr} is an extension of the
Standard Model (SM), in which  the gauge interactions of left-handed
and right-handed fundamental fermions are treated on equal basis.
There could be reflections of this extended structure in low-energy
phenomena due to possible mixings of the known particles with the new
particles predicted by the model, such as  deviations from the V-A
form of weak interactions, but nothing of that kind has turned up in
experiments so far.  The LRM is based on the gauge symmetry \ssu, and
there are hence two new weak bosons, $W_R$ and $Z_R$,  in addition to
the ones known in the SM. The  left-right symmetry, not present  in
the low energy world, is broken by a SU(2)$_R$ triplet Higgs field
$\Delta=(\Delta^{++},\Delta^+,\Delta^0)$. The only new fermions the
model predicts are the right-handed neutrinos.

The energy scale
$v_R=\vev{\Delta^0}$ of the breaking of the LRM  symmetry to the SM
symmetry, which also sets, up to coupling constants, the mass scale
of the  new weak bosons and right-handed neutrinos, is not given by
the theory itself. Its value can only be deduced from experimental
data. There exist several indirect bounds from processes where some
of the new particles appear as virtual states, as well as direct
discovery limits. According to these results it is not excluded that
the left-right symmetry manifests itself already at the scale of
$O(1)$ TeV. Hence, the search for direct evidences of the left-right
symmetry is a relevant issue to be considered in connection with the
Next Linear Collider (NLC).

In this note we shall investigate one particular reaction possible in
NLC, namely the production of the right-handed weak boson $W_2$ (we
denote this particle  $W_2$ instead of $W_R$ as there may be a slight
mixing between
$W_R$ and $W_L$) and a
 neutrino in the electron-photon collision (see also discussion in
\cite{aarre})):

\be e^-+\gamma\to W_2+N.
\label{reaction}\ee Here $N$ can be either a heavy right-handed
neutrino $\nu_R$ or, in the case of a large neutrino mixing, an
ordinary light neutrino $\nu_L$.
 So far,  the $ e^-\gamma$ collisions have been studied using  the
photon spectrum of  classical Bremsstrahlung.  In the linear collider
it will be possible to obtain high luminosity  photon beams by
backscattering intensive  laser pulses off the electron beam
\cite{ginz} without considerable losses in the  beam energy and with
very high polarizability and monochromaticity \cite{telnov}.

If the neutrino is light, the reaction (\ref{reaction}) would be the
energetically most favourable place to produce $W_2$. It has clean
signatures; if neutrino does not decay in the detector, the signal
will be two energetic jets and missing energy, or a single lepton and
missing energy. Using polarized beams will significantly reduce
backgrounds. The reaction would be useful in yielding valuable
information on both the gauge boson and neutrino masses.

Let us first discuss the existing bounds on the mass $M_{W_2}$ of the
new charged boson. In the Tevatron one has made a direct search of
$W_2$ in the channel $pp\to W_2\to eN$. The bound they give is
$M_{W_2}\gsim 652$ GeV \cite{mass}. The result is based on several
assumptions on the LRM: the quark-$W_2$ coupling has the SM strength,
the CKM matrices for the left-handed quarks and the right-handed
quarks are similar, the right-handed neutrino does not decay in the
detector but appears as missing $E_T$. If one relaxes the  first two
assumptions, the mass bound will be  weakened considerably, as was
pointed out in \cite{rizzoap95}. The third assumption is also
crucial; if the right-handed neutrino is heavy, with a mass of say
100 GeV or more, it will decay in the detector  into charged
particles (one possible channel could be
$\nu_R\to l+q\bar{q'}$) with no missing energy. For this case the
Tevatron search would be ineffective.

If the right-handed neutrino is heavier than $W_2$, the reaction
considered in the Tevatron search is forbidden altogether if there is
no substancial mixing between left- and right-handed neutrinos. In
this case
$W_2$ could decay only to non-leptonic states. Obviously, in the
Tevatron  hadronic signals are much more difficult to identify than
leptonic ones, and therefore the mass limit obtained from hadronic
decay channels is not particularly restrictive.

There are not very definite mass limits for the right-handed Majorana
neutrinos. At low energies they are assumed to be almost sterile;
they do couple to the ordinary $W$ and $Z$ bosons  due to their
mixing with the left-handed neutrinos \cite{gro}. This mixing is,
however, quite small, of the order of
$m_{l,q}/m_R$, where $m_{l,q}$ is a charged lepton or quark mass and
$m_R$ is the mass scale of the breaking of the left-right symmetry.
It is also possible that the ordinary weak bosons contain a small
fraction of $W_R$ and $Z_R$ states, which would allow the
right-handed neutrinos to couple to them, but also such mixings are
known to be small \cite{gun}.  Hence the laboratory mass limits for
the right-handed neutrinos are not that stringent.

Cosmological constraint on the mass of the right-handed neutrinos
derives from the condition that the  energy density associated with
them should not exceed the estimated total energy density of the
Universe. It was shown in
\cite{enq} that if the mixing between the left-handed and
right-handed sectors is on the level of 0.1, the mass of the
right-handed neutrino should exceed 30 GeV.

The conclusion from above is that there is still some uncertainty in
the mass bounds on the $W_2$ and the right-handed neutrinos. It was
argued in
\cite{rizzoap95}, that the lower limit for $M_{W_2}$ could be as low
as 300 GeV.

We wish to emphasize the usefulness of the reaction (\ref{reaction})
in studying the production and the mass relation of $W_2$ and the
right-handed neutrinos
$\nu_R$ (the right-handed electron neutrino if neutrinos of different
families do not mix).   This reaction has  advantages compared with
the corresponding processes in hadron colliders, such as Tevatron and
LHC. At NLC, good heavy flavour tagging would make the detection of
$W_2$ and $\nu_R$ efficient. Also, here the $W_2$ production is
insensitive to the unknown CKM matrix of the right-handed quarks. In
addition, the possibility of having highly polarized electron and
photon beams provides new probes for testing various properties of
the new particles
\cite{cuy}.

 There are two Feynman diagrams
 contributing at the tree level to  reaction (\ref{reaction}) (see
Fig. 1). The helicity amplitudes of the reaction for  general V,A
-interactions are given in ref. \cite{Martti}. In determining the
cross sections from the helicity amplitudes we have assumed 100\%
 polarized  electron and  photon beams. This is, of course, an
approximation, since   in practice the polarizations will never be
ideal and one has to employ a
 density matrix giving the polarization parameters of the beams.

  The mass dependence of the  total cross section of the process $
e^-_R\g\rightarrow W^-_2 N$ can be seen in Fig. 2, where we plot the
cross section as a function of
$ W^-_2$ mass for the center of mass energy
$ \sqrt{s_{ e\g} }=1.5$ TeV, expected to be possible to achieve in
the final stage of NLC, assuming the left- (Fig. 2 (I)) and
right-handedly (Fig. 2 (II)) polarized photon beams. The curves
denoted by $ a$ and $ b$ correspond to the neutrino masses
 $ M_N=300$ GeV and $ M_N=600$ GeV, respectively. The cross sections
are found to be reasonably large for almost the entire kinematically
allowed mass region, decreasing faster with $ M_{ W_2 }$ for the $
\tau_1=1$ photons. At low $ W_2$ masses the difference between $ a$
and $ b$ curve is small but for heavy $ W_2$ masses the cross section
depends strongly  on the neutrino mass. If $ M_N\leq M_{ W_2 },$
reaction (1) enables us to study heavier vector bosons than what is
possible  in the $ W^-_2$ pair production in
$ e^-e^+$ or $ e^-e^-$ collisions.

The reaction  would be even more useful in this respect
  if the mixing between the heavy and the light neutrino  is large
enough to give observable effects. In Fig. 3 we plot the cross
section of the  reaction
$ e^-_R\g\rightarrow W^-_2 \nu$ for different photon polarizations
assuming a vanishing  mass of $ \nu$ and the neutrino mixing angle of
$ \sin\theta_N=0.05.$   For this set of parameters the process should
be observable  up to $ W$-boson mass $ M_W=1.2$ TeV.

In order to see the relative importance
 of different polarization states we  present in Fig. 4 the angular
distributions  of the differential cross sections for all
combinations of polarizations. In Fig. 4 (I) we plot the
differential cross sections for the left-handedly  polarized photon
beam assuming the collision energy
$ \sqrt{s_{ e\g} }=1.5$ TeV, the gauge boson mass $ M_{ W_2 }=700$ GeV
and the neutrino mass $ M_N=300$ GeV.   In Fig. 4 (II) we plot the
same for the right-handedly polarized photon beam. As the figures
show,  the final states with a left-handed  neutrino  are clearly
suppressed. The main part of the  cross section is again coming from
the case where the photon and
$ W^-$ are polarized in the same way.  The other polarization
combinations are somewhat suppressed for the right-handedly polarized
photon beam. While in the SM {\em all \/} differential cross sections
are peaked in the backward direction,  in this case the distributions
are more flat for many polarization states  providing better
possibilities for detecting the anomalous photon coupling.

In conclusion, we have pointed out the usefulness of the electron
photon collisions in testing the left-right symmetry of electroweak
interactions at high energies through the reaction
$e^-\gamma\to W_2N$. If the right-handed neutrino is light, this
reaction  offers a much better discovery reach for $W_2$ than the
pair production in $e^+e^-$ or $e^-e^-$ collisions.

\bigskip

\noindent{\bf Acknowledgements.}  We are indebted to Aarre Pietilä,
Turku, for useful discussions. One of us (M.R.)
 expresses  his gratitude to Emil Aaltosen S\"a\"ati\"o, Wihurin
Rahasto  and Väisälän Rahasto for grants. This work has been
supported by the Academy of Finland.

\newpage
\section*{Figure captions}

\begin{description}

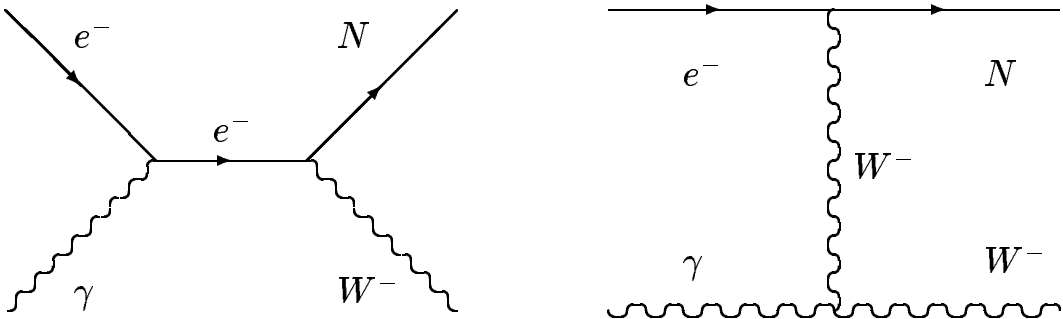
\item[Fig. 1.] Feynman diagrams for the  process $\protect
e^-\gamma\rightarrow W^-N$.
\vspace*{0.3cm}

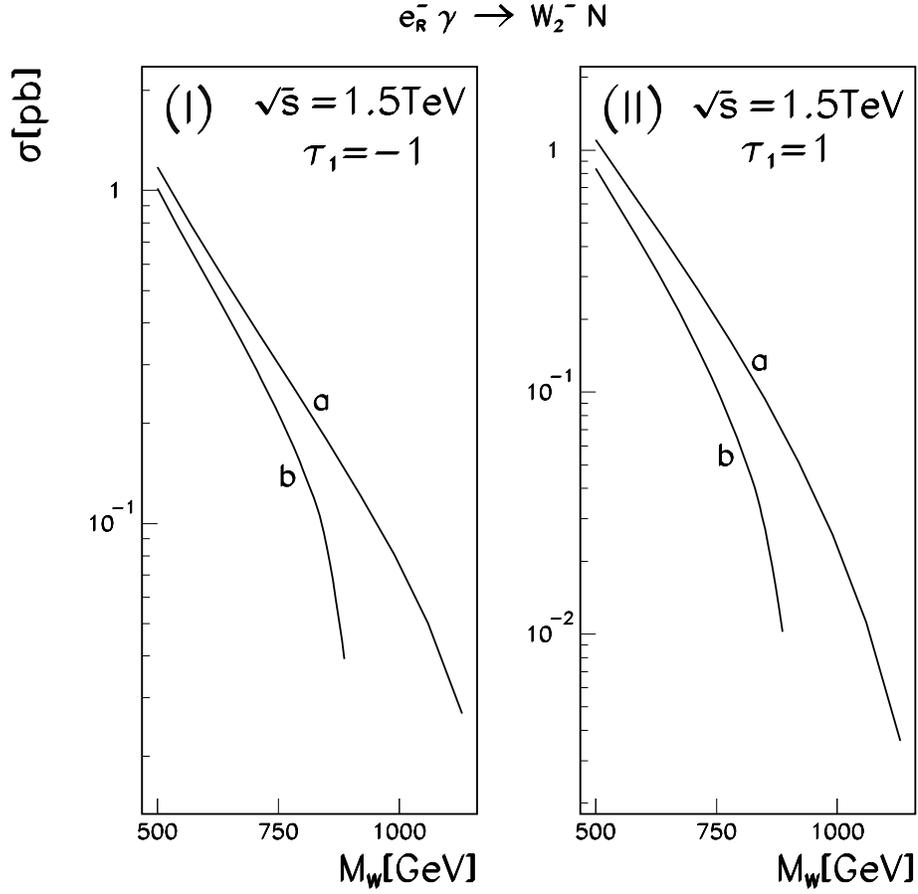
\item[Fig. 2.] The total cross section of the process
$\protect e_R^-\gamma\rightarrow W_{2}^-N $  as a function of heavy
gauge boson mass for the left- (figure (I)) and right-handedly
(figure (II)) polarized photon beams. The collision energy is taken
to be $\protect
\sqrt{s_{e\gamma}}=1.5$ TeV, and the mass of heavy neutrino $\protect
M_{N}=300$ GeV and $\protect M_{N}=600$  GeV for curves $a$ and $b,$
respectively.
\vspace*{0.3cm}

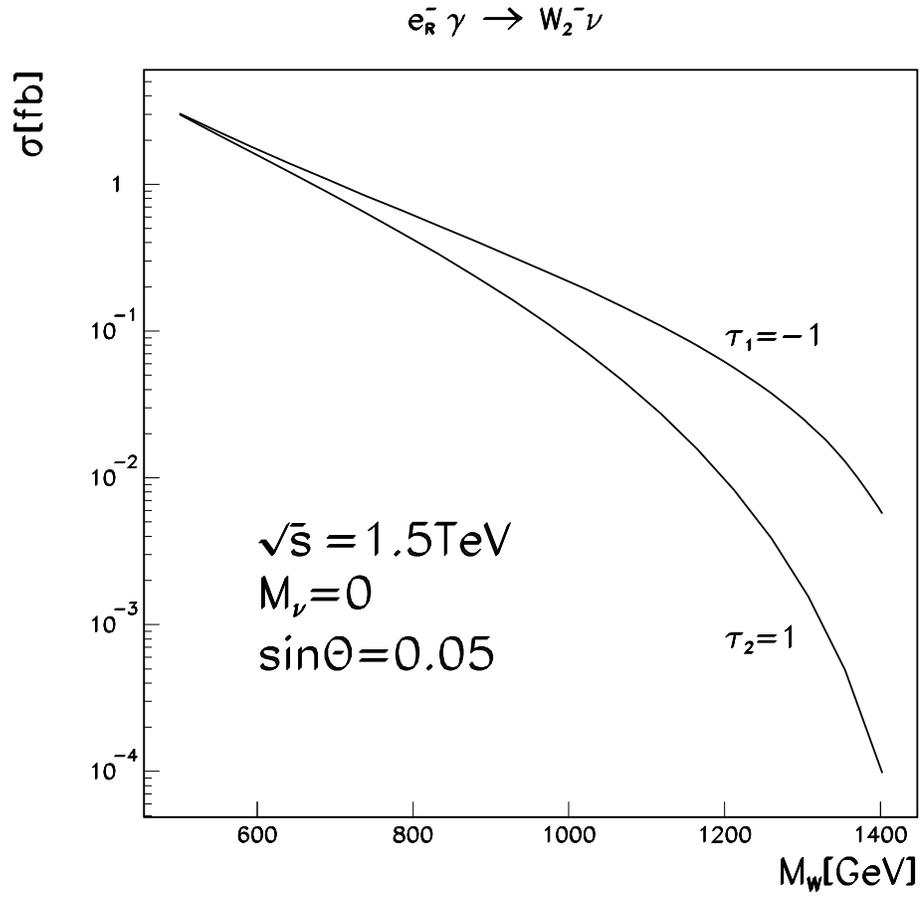
\item[Fig. 3.] The total cross section of the process
$\protect e_R^-\gamma\rightarrow W_{2}^-\nu $  as a function of heavy
gauge boson mass for the left- and right-handedly polarized photon
beams. The collision energy is taken to be $\protect
\sqrt{s_{e\gamma}}=1.5$ TeV and the neutrino mixing angle $\protect
\sin\theta=0.05.$
\vspace*{0.3cm}

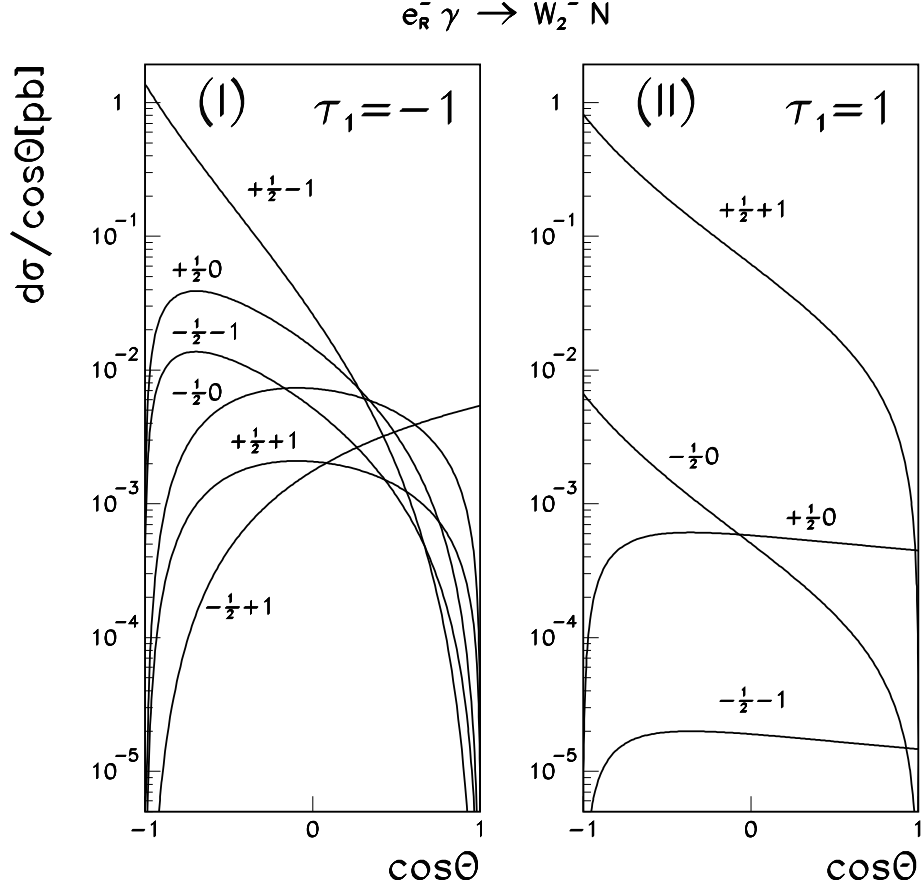
\item[Fig. 4.] The angular distributions of differential  cross
sections of various $\protect N W_2^-$ polarization states for  the
left- (figure (I)) and right-handedly (figure (II)) polarized  photon
beams in the case of  left-right model.  The collision energy is
taken to be $\protect
\sqrt{s_{e\gamma}}=1.5$ TeV,  the mass of right handed vector boson
$\protect M_W=700$ GeV and the mass of heavy neutrino $\protect
M_{N}=300$ GeV.  The number pairs in figure denote the helicity
states of neutrino and gauge boson, respectively.
\end{description}

\end{document}